\input harvmac 

\input amssym.def
\input amssym.tex

\def\Cop{\Bbb C}
\def\Zop{\Bbb Z}
\def\A{{\cal A}}

\lref\SenYV{
A.~Sen,
``Symmetries, conserved charges and (black) holes in two dimensional
string theory,''
arXiv:hep-th/0408064.
}

\lref\LarsenWC{
F.~Larsen, A.~Naqvi and S.~Terashima,
``Rolling tachyons and decaying branes,''
JHEP {\bf 0302}, 039 (2003)
[arXiv:hep-th/0212248].
}

\lref\GIR{D.~Gaiotto, N.~Itzhaki and L.~Rastelli,
``Closed strings as imaginary D-branes,''
Nucl.\ Phys.\ B {\bf 688}, 70 (2004) 
[arXiv:hep-th/0304192].
}

\lref\deBoerHD{
J.~de Boer, A.~Sinkovics, E.~Verlinde and J.~T.~Yee,
``String interactions in c = 1 matrix model,''
JHEP {\bf 0403}, 023 (2004)
[arXiv:hep-th/0312135].
}

\lref\SenZM{
A.~Sen,
``Rolling tachyon boundary state, conserved charges and two
dimensional string theory,''
JHEP {\bf 0405}, 076 (2004)
[arXiv:hep-th/0402157].
}

\lref\NelsonA{
I.~Affleck, W.~Hofstetter, D.~R.~Nelson and U.~Schollw\"ock,
``Non-Hermitian Luttinger liquids and flux line 
pinning in planar superconductors'', arXiv:cond-mat/0408478.
}

\lref\SenNU{
A.~Sen,
``Rolling tachyon,''
JHEP {\bf 0204}, 048 (2002)
[arXiv:hep-th/0203211].
}

\lref\SenIN{
A.~Sen,
``Tachyon matter,''
JHEP {\bf 0207}, 065 (2002)
[arXiv:hep-th/0203265].
}

\lref\CallanUB{
C.~G.~Callan, I.~R.~Klebanov, A.~W.~W.~Ludwig and J.~M.~Maldacena,
``Exact solution of a boundary conformal field theory,''
Nucl.\ Phys.\ B {\bf 422}, 417 (1994)
[arXiv:hep-th/9402113].
}

\lref\PolchinskiMY{
J.~Polchinski and L.~Thorlacius,
``Free fermion representation of a boundary conformal field theory,''
Phys.\ Rev.\ D {\bf 50}, 622 (1994)
[arXiv:hep-th/9404008].
}

\lref\GGone{
M.~B.~Green and M.~Gutperle,
``Symmetry breaking at enhanced symmetry points,''
Nucl.\ Phys.\  B {\bf 460}, 77 (1996) 
[arXiv:hep-th/9509171].
}

\lref\RSone{
A.~Recknagel and V.~Schomerus,
``Boundary deformation theory and moduli spaces of D-branes,''
Nucl.\ Phys.\ B {\bf 545}, 233 (1999) 
[arXiv:hep-th/9811237].
}

\lref\GaberdielXM{
M.~R.~Gaberdiel, A.~Recknagel and G.~M.~T.~Watts,
``The conformal boundary states for SU(2) at level 1,''
Nucl.\ Phys.\ B {\bf 626}, 344 (2002)
[arXiv:hep-th/0108102].
}

\lref\GaberdielZQ{
M.~R.~Gaberdiel and A.~Recknagel,
``Conformal boundary states for free bosons and fermions,''
JHEP {\bf 0111}, 016 (2001)
[arXiv:hep-th/0108238].
}

\lref\GutperleAI{
M.~Gutperle and A.~Strominger,
``Spacelike branes,''
JHEP {\bf 0204}, 018 (2002)
[arXiv:hep-th/0202210].
}

\lref\GutperleXF{
M.~Gutperle and A.~Strominger,
``Timelike boundary Liouville theory,''
Phys.\ Rev.\ D {\bf 67}, 126002 (2003)
[arXiv:hep-th/0301038].
}

\lref\LambertZR{
N.~Lambert, H.~Liu and J.~Maldacena,
``Closed strings from decaying D-branes,''
arXiv:hep-th/0303139.
}

\lref\McGreevyKB{
J.~McGreevy and H.~Verlinde,
``Strings from tachyons: The c = 1 matrix reloaded,''
JHEP {\bf 0312}, 054 (2003)
[arXiv:hep-th/0304224].
}

\lref\KlebanovKM{
I.~R.~Klebanov, J.~Maldacena and N.~Seiberg,
``D-brane decay in two-dimensional string theory,''
JHEP {\bf 0307}, 045 (2003)
[arXiv:hep-th/0305159].
}

\lref\StromingerPC{
A.~Strominger,
``Open string creation by S-branes,''
arXiv:hep-th/0209090.
}

\lref\FredenhagenUT{
S.~Fredenhagen and V.~Schomerus,
``On minisuperspace models of S-branes,''
JHEP {\bf 0312}, 003 (2003)
[arXiv:hep-th/0308205].
}

\lref\FStwo{
S.~Fredenhagen and V.~Schomerus,
``Boundary Liouville theory at $c=1$,''
arXiv:hep-th/0409256.
}

\lref\SenYV{
A.~Sen,
``Symmetries, conserved charges and (black) holes in two dimensional string
theory,''
arXiv:hep-th/0408064.
}

\lref\OkudaYD{
T.~Okuda and S.~Sugimoto,
``Coupling of rolling tachyon to closed strings,''
Nucl.\ Phys.\ B {\bf 647}, 101 (2002)
[arXiv:hep-th/0208196].
}

\lref\OkuyamaJK{
K.~Okuyama,
``Comments on half S-branes,''
JHEP {\bf 0309}, 053 (2003)
[arXiv:hep-th/0308172].
}

\lref\BG{C.~Bachas and M.~R.~Gaberdiel, 
``World-sheet duality for D-branes with travelling waves,''
JHEP {\bf  0403}, 015 (2004) 
[arXiv:hep-th/0310017].
}

\lref\vijay{
V.~Balasubramanian, E.~Keski-Vakkuri, P.~Kraus and
A.~Naqvi,
``String scattering from decaying branes,''
arXiv:hep-th/0404039.
}

\lref\StromingerTT{
A.~Strominger and T.~Takayanagi,
``Correlators in timelike bulk Liouville theory,''
Adv.\ Theor.\ Math.\ Phys.\ {\bf 7}, 369 (2003)
[arXiv:hep-th/0303221].
}

\lref\ConstableL{
N.~R.~Constable and F.~Larsen, 
``The rolling tachyon as a matrix model,'' JHEP
{\bf 0306}, 017 (2003) [arXiv:hep-th/0305177].
}

\lref\Schomerus{
V.~Schomerus,
``Rolling tachyons from Liouville theory,''
JHEP {\bf 0311}, 043 (2003) [arXiv:hep-th/0306026].
}

\lref\KLMS{
J.~L.~Karczmarek, H.~Liu, J.~Maldacena and A.~Strominger,
``UV finite brane decay,''
JHEP {\bf 0311}, 042 (2003) [arXiv:hep-th/0306132].
}

\Title {\vbox{ \baselineskip12pt
\hbox{hep-th/0410098}\hbox{UCLA/04/TEP-41}
}} {\vbox{ \centerline{Remarks on the rolling tachyon BCFT}  }}

\centerline{ Matthias R. 
Gaberdiel$^a$\footnote{$^\ast$}{\tt email:
gaberdiel@itp.phys.ethz.ch} and 
Michael Gutperle$^b$\footnote{$^\dagger$}{\tt email:
gutperle@physics.ucla.edu}}

\bigskip
\centerline{ $^a$ Theoretische Physik, ETH-H\"onggerberg, CH-8093
Z\"urich, Switzerland} 
\medskip
 
\centerline{ $^b$ Department of Physics and Astronomy, UCLA,
Los Angeles,  CA 90095, USA}
\medskip

\vskip .3in \centerline{\bf Abstract}

\smallskip\noindent
It is shown how the boundary correlators of the Euclidean theory
corresponding to the rolling tachyon solution can be calculated
directly from Sen's boundary state. The resulting formulae reproduce
precisely the expected perturbative open string answer. We also
determine the open string spectrum and comment on the implications of
our results for the timelike theory.     

\vfill

\smallskip
\Date{October 2004}

\newsec{Introduction}
The study of time dependent string theory solutions has recently been
a very active area of research. The creation and decay of an unstable
D-brane constitutes one of the simplest examples of a time dependent 
background. Sen \refs{\SenIN,\SenNU} showed that the rolling of the
open string tachyon in time can be described by the boundary
deformation
\eqn\one{S= -{1\over 2\pi} 
\int_\Sigma d\tau d\sigma\,  \partial_a X^0 \partial^a X^0 +
{\lambda } \int_{\partial \Sigma} 
d\tau \cosh\big(\sqrt{2}X^0(\tau)\big) \,.}
The boundary deformation in \one\ can be interpreted as a non-zero
background for the tachyon field 
$T(X^0)= \lambda \cosh(\sqrt{2}X^0)$. At early and late times, 
$X^0\to \pm \infty$, the tachyon approaches its  
minimum $T\to \infty$ where the unstable brane has disappeared. Hence 
\one\ gives a concrete realisation of a `S(pacelike)-brane'
\GutperleAI. 

Sen showed that \one\ defines an exact boundary conformal field theory
(BCFT) by relating the timelike theory to a spacelike theory by means
of the analytic continuation $X^0\mapsto i X$. The Euclidean boundary
theory  
\eqn\two{S= 
{1\over 2\pi} \int_\Sigma d\tau d\sigma\, \partial_a X \partial^a X + 
{\lambda } \int_{\partial \Sigma} d\tau \cos\big(\sqrt{2}X(\tau)\big) }
is then conformal since the boundary perturbation is exactly marginal,
as follows from the underlying $SU(2)$ symmetry of a boson
compactified on a circle at the self dual radius
\refs{\CallanUB,\PolchinskiMY}. 
In conformal field theory D-branes are
described by boundary states; for a non-compact boson associated with
the deformation \two\ the relevant boundary state is 
\refs{\CallanUB,\GGone,\RSone,\GaberdielXM,\GaberdielZQ}
\eqn\three{|\!|B\rangle\!\rangle = {1\over 2^{1/4}} 
\sum_{j,m}
D^j_{-m,m}\pmatrix{ \cos\pi\lambda &-\sin\pi \lambda \cr 
\sin\pi \lambda & \cos\pi \lambda}
|j, m,m\rangle\!\rangle \,,}
where $D^j_{-m,m}(g)$ is the $(-m,m)$-matrix element of $g\in SU(2)$ 
in the representation with spin $j$ (which in our conventions is a
non-negative half-integer), and  $|j, m,m\rangle\!\rangle$ is the
Virasoro Ishibashi state (labelled by $j$) in the sector with 
$(p_L,p_R)=2 \sqrt{2}(m,m)$. (Note that $m$ is here also 
half-integer. An explicit formula for the matrix elements
$D^j_{m,n}(g)$ can for example be found in \GaberdielXM.) 

\noindent A different rolling tachyon, called the 'half S-brane', was
introduced in \refs{\GutperleXF,\LarsenWC} 
\eqn\four{S= -{1\over 2\pi} 
\int_\Sigma d\tau d\sigma\,  \partial_a X^0 \partial^a X^0 + 
{\lambda } \int_{\partial \Sigma} d\tau e^{\sqrt{2}X^0} \,.}
This tachyon background describes the decay of an unstable brane since
the tachyon approaches its maximum (where the tachyon has not
condensed and the D-brane is present) at early times, while it obtains
its minimum at late times. An analytic continuation to the spacelike
theory produces now   
\eqn\five{S= {1\over 2\pi} 
\int_\Sigma d\tau d\sigma\, \partial_a X \partial^a X +
{\lambda } \int_{\partial \Sigma} d\tau e^{i\sqrt{2}X} \,.}
The boundary state associated to \five\ is given by
\eqn\six{|\!| B\rangle\!\rangle=
\sum_{j=0,{1\over 2},\cdots} \;
\sum_{m\geq 0}^j 
\pmatrix{j+m\cr 2m} (-\lambda )^{2m} |j,m,m\rangle\!\rangle
\,.}
This is the boundary state \three\ corresponding to the group element 
$g_\lambda\in {\rm SL}(2,\Cop)$ \refs{\GaberdielXM,\GaberdielZQ},
where 
\eqn\rsixa{
g_\lambda = \pmatrix{1 & -\lambda \cr 0 & 1} \,.}
The exact conformal field theory description of the rolling tachyon
in terms of its boundary state has been used to investigate many
aspects of tachyon condensation and brane decay, in particular, the
production of closed strings in the decay \refs{\LambertZR,\GIR},
tachyon matter \SenIN, two dimensional string theory
\refs{\McGreevyKB,\KlebanovKM} and comparison to minisuperspace
calculations \refs{\StromingerPC,\FredenhagenUT}.

On the other hand, various aspects of this boundary conformal field
theory still remain mysterious. The analytic continuation from the
spacelike theory to the timelike theory is still not satisfactorily
understood, since it is not clear at which stage in the calculation
the analytic continuation should be performed. The timelike boundary
states contain components which diverge in the $X^0\to \infty$ limit
\OkudaYD. Although the divergent components do not contribute in some
calculations (like the bulk one point functions responsible for the
closed string creation), they nevertheless play an important role for
conserved charges associated with branes \refs{\SenYV,\SenZM} and the
open string interpretation of the boundary conformal field theory
\OkuyamaJK.  

At an even simpler level, various string amplitudes (of the Euclidean
or the timelike theory) have only been calculated rather indirectly,
in particular, by using the relation of the theory to a particular
limit of Liouville theory (see for example
\refs{\StromingerPC,\GutperleXF,\StromingerTT,\ConstableL,\Schomerus,\FStwo}). 
However, given the relative simplicity of this model, it should be
possible to derive these amplitudes directly from the conformal field
theory description of the D-brane in terms of its boundary state.  
In this paper we want to explain how this can be done, at least for
the case of the boundary correlators of the Euclidean theory. More
precisely, we shall explain how these correlation functions can be
obtained directly from the 
Euclidean boundary state \six, following ideas of \BG. The resulting
expressions reproduce precisely the expected open string 
expansion\foot{While this is certainly the expected answer, it is
unclear to us how to derive it more abstractly since the `marginal
field' by which the boundary state is perturbed is not actually
present in the theory at infinite radius. In any case, the detailed
relation between these two calculations is instructive, and may be
useful in other contexts.
For example, our analysis gives rise to a prescription for how to
evaluate these formally divergent integrals.}
that was, for example, used as the starting point in \vijay. We also
explain how the techniques we describe should allow one to determine
other string amplitudes directly from the boundary state. 

In this paper we focus on the case of a single free boson that
can be described in terms of SU(2) boundary states. The
calculational techniques we describe can however also be applied to
boundary states that are associated to groups other than SU(2).

\medskip

The paper is organised as follows. In section 2 we introduce our
conventions and explain the calculation of the boundary $n$-point
function of the Euclidean theory in detail. We also mention how the
open string spectrum can be derived from this boundary state, and how
these results fit together. Finally we comment in section 3 on the
continuation to the timelike case. We have included an appendix in
which some of the more technical parts of our calculations are
explained.

\newsec{Boundary correlators and the cylinder}

In this section we want to calculate the boundary correlators of the
Euclidean rolling tachyon solution at $c=1$. We begin by describing
our conventions in detail.

\subsec{Conventions}
The spacelike closed string field $X(\sigma,\tau)$ has the mode
expansion  (in the following we shall always set $\alpha'=1/2$)
\eqn\rone{
X(\sigma,\tau) = x_0 + {1\over 2}\, p \tau + {i \over 2} \sum_{n\ne 0} 
{1\over n} e^{-i n \tau} \, 
\left(\alpha_n e^{-in\sigma} + \tilde\alpha_n e^{in\sigma} \right)
\,.}
The canonical commutation relations are 
\eqn\rtwo{
{}[\alpha_m,\alpha_n] = [\tilde\alpha_m,\tilde\alpha_n] = 
m\, \delta_{m,-n}\,, \qquad
[x,p] = i \,.}
As usual, the boundary state will be inserted at $\tau=0$; for the
following it is therefore useful to define the positive and negative
parts of $X$ at $\tau=0$:
\eqn\rthree{\eqalign{
X_>(\sigma) & =  {i \over 2} \sum_{n > 0} 
{1\over n} \, 
\left(\alpha_n e^{-in\sigma} + \tilde\alpha_n e^{in\sigma} \right)\,,
\cr
X_<(\sigma) & =  {i \over 2} \sum_{n < 0} 
{1\over n} \, 
\left(\alpha_n e^{-in\sigma} + \tilde\alpha_n e^{in\sigma} \right)
\,.}}
The commutator of the positive and negative modes is 
\eqn\rfour{\eqalign{
[X_>(\sigma_1),X_<(\sigma_2)] & =  {1\over 4} \sum_{n>0}
{1\over n} \left(e^{-in(\sigma_1-\sigma_2)} +
e^{in(\sigma_1-\sigma_2)} \right) \cr
& =  - {1\over 4} \, \log \left(1 - e^{-i(\sigma_1-\sigma_2)} \right) 
- {1\over 4}\, \log \left(1 - e^{i(\sigma_1-\sigma_2)} \right)  \cr
& =  - {1\over 4} \, \log 
\left[4\sin^2 \left({\sigma_1-\sigma_2\over 2} \right) \right] \,.}}
If the spacelike boson is compactified on a circle of radius $R$, then
the left- and right-moving momenta are quantised as 
\eqn\rfive{
(p_L,p_R) = \left({\hat{m}\over R} + 2 \hat{n} R, 
{\hat{m}\over R} - 2 \hat{n} R \right) \,,}
where $\hat{m},\hat{n}\in\Zop$. The self-dual radius is
$R_{sd}=1/\sqrt{2}$; at this point, the momenta are simply
\eqn\rsix{
(p_L,p_R) = \sqrt{2} \left(\hat{m}+\hat{n},
\hat{m}-\hat{n}\right)\,.}
At the self-dual point, the symmetry is enhanced from u(1) to su(2);
the relevant (left-moving) operators are (up to cocycle factors) given
as   
\eqn\rseven{
J^\pm(z) = :\exp\left(\pm i 2 \, \sqrt{2} X_L(z)\right): \,, 
\qquad 
J^3(z) = \sqrt{2} i \partial_z \, X_L(z) \,,}
where $z=e^{i(\tau+\sigma)}$, and  
\eqn\reight{
X_L(z) =  {1\over 2}\, x_0  - {i\over 4}\, p_L \, \log (z)  
+ {i \over 2} \sum_{n\ne 0} {1\over n}\, \alpha_n\, z^{-n} \,.}
[Obviously there is a similar extended right-moving symmetry,
generated  by corresponding formulae in which $X_L(z)$ is replaced by 
$X_R(\bar{z}$) with $\bar{z} = e^{i(\tau-\sigma)}$.]
The normal ordering in \rseven\ is the usual chiral normal ordering,
where positive modes are moved to the right of negative
modes. The corresponding
modes satisfy the standard su$(2)_1$ commutation relations,
\eqn\rnine{\eqalign{
{}[J^3_m,J^\pm_n] & =  \pm J^\pm_{m+n} \cr
{}[J^+_m,J^-_n] & =  2\, J^3_{m+n} + m\,\delta_{m,-n} \cr
{}[J^3_m,J^3_n] & = {1\over 2}\, m\, \delta_{m,-n} \,. }}
For future reference we observe that $J^3_n={1\over\sqrt{2}}\alpha_n$.

\subsec{Disc amplitudes}

We are interested in calculating various disk diagrams involving the
boundary state corresponding to the group element 
$g_\lambda\in {\rm SL}(2,\Cop)$ \rsixa. As is explained in 
\refs{\CallanUB,\GGone,\RSone,\GaberdielXM,\GaberdielZQ}, we can write
this boundary state also as 
\eqn\szero{
|\!| B \rangle\!\rangle_\infty = P_\infty
\, \exp\left(-\lambda J^+_0 \right) \, |\!| e \rangle\!\rangle \,,}
where $J^+_0$ is the (left-moving) chiral mode, 
$|\!| e \rangle\!\rangle$ denotes the boundary state corresponding to
the identity state for the SU(2) theory at the self-dual radius, and 
$P_\infty$ is the projector onto states for which $p_L=p_R$. We also
write  
\eqn\szeroa{
|\!|N\rangle\!\rangle \equiv 
|\!| e \rangle\!\rangle_\infty }
for the usual Neumann boundary state (in the uncompactified
situation). 

We are interested in calculating the $n$-point  function of $n$
boundary vertex operators, using the above boundary state description
of the D-brane. Adopting the method of \BG\ to this context,
one can calculate this as follows. We define the {\it Neumann normal
ordering} of an open  string vertex operator at $\tau=0$ as  
\eqn\sone{
: e^{i\omega X(\sigma)} :_N \equiv
e^{2i\omega X_<(\sigma)} \, e^{i\omega x_0} \, 
e^{i\omega (X_>(\sigma) - X_<(\sigma))} \,.}
Actually, as will be shown in the appendix, this normal ordering
prescription corresponds precisely to considering the leading
term in the bulk-boundary OPE that describes the limit in which 
a bulk operator approaches the boundary.\foot{We thank Andreas
Recknagel for helping us understand this issue.} In particular, this 
observation therefore determines the normalisation of our boundary
vertex operators: the normalisation of the boundary vertex operators
is such that the above bulk-boundary OPE coefficient is unity.  

The motivation for the normal ordering prescription \sone\ is that the
exponent on the right annihilates the Neumann boundary state
$|\!|N\rangle\!\rangle$,  
\eqn\stwo{
(X_>(\sigma) - X_<(\sigma)) \, |\!|N\rangle\!\rangle = 
{i\over 2} \sum_{n>0} {1\over n} \left[
\left(\alpha_n + \tilde\alpha_{-n} \right) e^{-in\sigma} 
+ \left(\tilde\alpha_n + \alpha_{-n} \right) e^{in\sigma} \right] 
\, |\!|N\rangle\!\rangle = 0 \,.}
In particular, this implies that the one-point function 
\eqn\sthree{
\langle 0 | \, : e^{i\omega X(\sigma)} :_N  \, 
 |\!|N\rangle\!\rangle  = \langle 0 | \,
e^{i\omega x_0}\, |\!|N\rangle\!\rangle}
vanishes unless $\omega=0$. Furthermore, the higher point
functions
\eqn\sfour{
\langle 0 | \, : e^{i\omega_1 X(\sigma_1)} :_N  \, \cdots \, 
: e^{i\omega_n X(\sigma_n)} :_N \, |\!|N\rangle\!\rangle}
converge, and agree in fact with the boundary correlators for the
corresponding operators. For the two point function  one easily shows that
\eqn\sfive{\eqalign{
\langle 0 | \, : e^{i\omega_1 X(\sigma_1)} :_N  \, 
: e^{i\omega_2 X(\sigma_2)} :_N \, |\!|N\rangle\!\rangle & =  
\langle 0 | \, e^{i (\omega_1+\omega_2) x_0} \, 
e^{i\omega_1 (X_>(\sigma_1) - X_<(\sigma_1))} 
e^{2i\omega_2 X_<(\sigma_2)} \,  |\!|N\rangle\!\rangle  \cr
& =  \left[ 4 \sin^2 \left({\sigma_1-\sigma_2 \over 2} \right) 
\right]^{{\omega_1 \omega_2 \over 2}}\, \langle 0 | \, 
e^{i (\omega_1+\omega_2) x_0} \, |\!|N\rangle\!\rangle  \cr
& =  \delta(\omega_1+\omega_2) \, 
\left[ 4 \sin^2 \left({\sigma_1-\sigma_2 \over 2} \right) 
\right]^{- {\omega_1^2 \over 2}}\,, }}
where we have used the Baker-Campbell-Hausdorff (BCH) formula,
\eqn\bch{
e^{B+C} = e^B \, e^{\half [C,B]} \, e^{C} \,,}
that holds provided that $[B,C]$ commutes with both $B$ and $C$, 
as well as \rfour.

The boundary correlation functions for the boundary state with the
boundary deformation \five\ turned on,
can be calculated by replacing  the Neumann boundary state
$|\!|N\rangle\!\rangle$  with \szero. [As is also shown in the
appendix, the above interpretation of this calculation as a limit of a
bulk calculation is also valid in this context; in particular, our
boundary vertex operators are therefore still normalised so that this
bulk-boundary OPE coefficient is unity.] Thus the two-point
function of such boundary vertex operators should be  
\eqn\ssix{
\A_2 = \langle 0 | \, : e^{i\omega_1 X(\sigma_1)} :_N  \, 
: e^{i\omega_2 X(\sigma_2)} :_N  \, 
|\!|B\rangle\!\rangle  \,.}
Performing the same calculation as above, one finds that $\A_2$ equals 
\eqn\sseven{
\A_2 = \left[ 4 \sin^2 \left({\sigma_1-\sigma_2 \over 2} \right) 
\right]^{{\omega_1 \omega_2 \over 2}}\, 
\langle 0 | \, e^{i (\omega_1+\omega_2) x_0} \, 
e^{i \omega_1 (X_>(\sigma_1) - X_<(\sigma_1)) + 
i \omega_2 (X_>(\sigma_2) - X_<(\sigma_2)) } \, 
|\!|B\rangle\!\rangle  \,.}
The exponential on the right hand side of \sseven\ can be written as
\eqn\seight{
 \exp\left[i \omega_1 (X_>(\sigma_1) - X_<(\sigma_1)) + 
i \omega_2 (X_>(\sigma_2) - X_<(\sigma_2)) \right] \equiv
\exp\left(Y\right)\,,} 
where
\eqn\snine{
Y = -{1 \over 2} \sum_{n\ne 0} {1\over |n|} \, 
(\alpha_n + \tilde\alpha_{-n} ) \gamma_n \,,}
and $\gamma_n$ is given as 
\eqn\sten{
\gamma_n = \omega_1\, e^{-in\sigma_1} + \omega_2\, e^{-in\sigma_2}
\,.} 
Now we observe that 
\eqn\seleven{
e^{Y} \, |\!|B\rangle\!\rangle  = 
e^{ Y} \, P_\infty \, e^{-\lambda J^+_0} \, |\!| e \rangle\!\rangle 
=   P_\infty \,
e^{ Y}\, e^{-\lambda J^+_0} \, |\!| e \rangle\!\rangle 
=  P_\infty \,
e^Z \, |\!| e \rangle\!\rangle \,, }
where we have used that $P_\infty$ commutes with $e^{Y}$ (since $Y$
does not involve any zero modes), and where
\eqn\stwelve{
Z = - \lambda \, \sum_{k=0}^{\infty} {1\over k!} \, 
{\rm Ad}^k (Y) J^+_0 \,.}
Here ${\rm Ad}^k (Y) J^+_0$ is the $k$-fold commutator
$[Y,[Y,\cdots,[Y,Z]\cdots ]]$ and we have used that
$Y|\!|e\rangle\!\rangle=0$ and that the commutator $[Y,J^+_0]$ is
a sum of terms involving only the modes $J^+_l$. Using that
$J^3_n = {1\over \sqrt{2}}\alpha_n$, as well as the SU(2) commutation
relations, one finds that 
\eqn\sthirteen{
Z = - \lambda \sum_{k=0}^{\infty} {1\over k!} 
\left(-{1\over \sqrt{2}}\right)^k
\sum_{n_1,\ldots,n_k} 
{\gamma_{n_1} \over |n_1|} \cdots {\gamma_{n_k} \over |n_k|} \,
J^+_{n_1+\cdots +n_k} \,,}
where the sum runs over all $n_i\ne 0$. The total amplitude is then 
\eqn\sfourteen{\eqalign{
{\cal A}_2 & =  \left[ 4 \sin^2 \left({\sigma_1-\sigma_2 \over 2} \right) 
\right]^{{\omega_1 \omega_2 \over 2}}\, 
\langle 0 | \, e^{i (\omega_1+\omega_2) x_0} \,
P_\infty \, e^Z \, |\!|e\rangle\!\rangle  \cr
& = \left[ 4 \sin^2 \left({\sigma_1-\sigma_2 \over 2} \right) 
\right]^{{\omega_1 \omega_2 \over 2}}\, 
\langle 0 | \, e^{i (\omega_1+\omega_2) x_0} \,
e^Z \, |\!|e\rangle\!\rangle \,,}}
since the state on the left is in the space onto which $P_\infty$
projects. For each $k$ the coefficient of $J^+_m$ in $Z$ is 
proportional to 
\eqn\ssixteen{\eqalign{
D_m & = \sum_{n_1,\ldots,n_k} 
{\gamma_{n_1} \over |n_1|} \cdots {\gamma_{n_k} \over |n_k|} \,
\delta_{n_1+\cdots + n_k,m}  \cr
& = 
\int_0^{2\pi} {dt\over 2\pi}\, 
\sum_{n_1,\ldots,n_k} e^{- i t (n_1+\cdots + n_k-m)} 
{\gamma_{n_1} \over |n_1|} \cdots {\gamma_{n_k} \over |n_k|} \,
\cr
& = 
\int_0^{2\pi} {dt\over 2\pi}\, e^{ i t m}
\left(\sum_{n\ne 0} {\gamma_n \over |n|} e^{- i t n} \right)^k
\,.}}
With $\gamma_n$ given by \sten\ the above sum can be performed explicitly,
\eqn\sseventeen{\eqalign{
\sum_{n\ne 0} {\gamma_n \over |n|} e^{- i t n} & = 
- \omega_1 \log\left(1 - e^{-i(\sigma_1 +  t)} \right)
- \omega_1 \log\left(1 - e^{i(\sigma_1 +  t)} \right) \cr
&  \qquad 
- \omega_2 \log\left(1 - e^{-i(\sigma_2 +  t)} \right)
- \omega_2 \log\left(1 - e^{i(\sigma_2 +   t)} \right) \,.}}
Thus the $J^+_m$ component of $Z$ becomes
\eqn\seighteen{\eqalign{
Z_m & =  - \lambda \,J^+_m \,  \int_0^{2\pi} {dt\over 2\pi}\, e^{ itm}
\exp\left( - {1\over \sqrt{2}} \, 
\sum_{n\ne 0} {\gamma_n \over |n|} e^{-i t n} \right) \cr
& =  
- \lambda \,J^+_m \,  \int_0^{2\pi} {dt\over 2\pi}\, e^{ itm}
\left(1 - e^{-i(\sigma_1 +  t)} \right)^{{\omega_1\over \sqrt{2}}} \,
\left(1 - e^{i(\sigma_1 +  t)} \right)^{{\omega_1\over \sqrt{2}}} \cr
&  \qquad \qquad \qquad \qquad \times
\left(1 - e^{-i(\sigma_2 +  t)} \right)^{{\omega_2\over \sqrt{2}}} \,
\left(1 - e^{i(\sigma_2 +  t)} \right)^{{\omega_2\over \sqrt{2}}}  \cr
& = 
- \lambda \,J^+_m \,  \int_0^{2\pi} {dt\over 2\pi}\, e^{ i tm}
\left( 4 \sin^2\left({\sigma_1 +  t \over 2} \right)
\right)^{\omega_1\over \sqrt{2}} \, 
\left( 4 \sin^2\left({\sigma_2 +  t \over 2} \right)
\right)^{\omega_2\over \sqrt{2}} \,.}}
Summing over all $m$ we therefore obtain
\eqn\snineteen{
Z = - \lambda \int_0^{2\pi} {dt\over 2\pi}\, J^+(e^{- i t})
\left( 4 \sin^2\left({\sigma_1 +  t \over 2} \right)
\right)^{\omega_1\over \sqrt{2}} \, 
\left( 4 \sin^2\left({\sigma_2 +  t \over 2} \right)
\right)^{\omega_2\over \sqrt{2}} \,,}
where 
\eqn\stwenty{
J^+(e^{- i t}) = \sum_{m\in\Zop} J^+_m e^{ i tm} \,.}
Finally we use that
\eqn\szwischen{
\langle 0|\, e^{i(\omega_1+\omega_2)x_0}\, 
J^+(e^{-it_1})\cdots J^+(e^{-it_m}) \, |\!| e \rangle\!\rangle 
= \delta(\omega_1+\omega_2 + \sqrt{2} m )\, 
\prod_{r<s}^m  \Big[ 4\sin^2\Big({t_r- t_s \over 2} \Big)\Big]\,,}
as follows from \rseven, performing a similar calculation as in 
\sfive. [The only minor difference is that now $\alpha_n$ with $n>0$
does not annihilate the boundary state $|\!|e\rangle\!\rangle$, but
rather $\alpha_n |\!|e\rangle\!\rangle = - \tilde{\alpha}_{-n}
|\!|e\rangle\!\rangle$. This right-moving creation mode then
annihilates the in-vacuum to the left.]  

Putting everything together, we therefore find that the two-point
function can be evaluated as  
\eqn\finaltwo{\eqalign{\A_2 &=  
\left[ 4 \sin^2 \left({\sigma_1-\sigma_2\over 2}  \right) 
\right]^{{\omega_1 \omega_2 \over 2}} \sum_{m=0}^{\infty}
{(-\lambda)^m\over m!}
\delta(\omega_1+\omega_2+\sqrt{2}m) \cr
&\quad \times
\prod_{l=1}^m  \int {dt_l\over 2\pi} 
\prod_{k=1}^2 \left( 4 \sin^2\left({\sigma_k +  t_l \over 2} \right)
\right)^{\omega_k\over \sqrt{2}} 
\prod_{r<s}^m  \Big[ 4\sin^2\Big({t_r- t_s \over 2} \Big)\Big]\,.}}
For a given $\omega_1$ and $\omega_2$, at most one term in the sum
over $m$ contributes.
\medskip

The calculation of the two point function can easily be generalised to
an $n$-point boundary correlator
\eqn\ssix{\eqalign{
\A_n &= \langle 0 | \, : e^{i\omega_1 X(\sigma_1)} :_N  \, 
: e^{i\omega_2 X(\sigma_2)} :_N  \cdots : e^{i\omega_n X(\sigma_n)} :_N \, 
|\!|B\rangle\!\rangle  \cr
& = \prod_{i<j}\left[ 4 \sin^2 \left({\sigma_i-\sigma_j \over 2} \right) 
\right]^{{\omega_i \omega_j \over 2}}\, 
\langle 0 | \, e^{i (\omega_1+\omega_2+\cdots +\omega_n) x_0} \,
e^Z \, |\!|e\rangle\!\rangle \,,}}
where $Z$ is now given by
\eqn\zncase{Z = - \lambda \int_0^{2\pi} 
{dt\over 2\pi}\, J^+(e^{-i t}) \prod_{i=1}^n 
\left( 4 \sin^2\left({\sigma_i +  t \over 2} \right)
\right)^{\omega_i\over \sqrt{2}}\,.}
The $n$-point function \ssix\ can thus be evaluated as a perturbation
series in $\lambda$,
\eqn\aperb{\eqalign{\A_n &= 
\prod_{i<j}\left[ 4 \sin^2 \left({\sigma_i-\sigma_j \over 2} \right) 
\right]^{{\omega_i \omega_j \over 2} }\sum_{m=0}^{\infty}
{(-\lambda)^m\over m!}
\prod_{l=1}^m \int {dt_l\over 2\pi}  
\prod_{k=1}^n \left( 4 \sin^2\left({\sigma_k +  t_l \over 2} \right)
\right)^{\omega_k\over \sqrt{2}}\cr
& \qquad\qquad \times \langle 0|\, e^{i(\omega_1+\cdots+\omega_n)x_0}
J^+(e^{- i t_1})\cdots  J^+(e^{- i t_m})\, |\!|e\rangle\!\rangle \,.}}
Using \szwischen\ we then obtain
\eqn\anper{\eqalign{\A_n &=  
\prod_{i<j}\left[ 4 \sin^2 \left({\sigma_i-\sigma_j
\over 2}
 \right) 
\right]^{{\omega_i \omega_j \over 2}} \sum_{m=0}^{\infty} 
{(-\lambda)^m\over m!}
\delta(\omega_1+\omega_2+\cdots+\omega_n +\sqrt{2}m) \cr
&\qquad\qquad \times
\prod_{l=1}^m  \int {dt_l\over 2\pi} 
\prod_{k=1}^n \left( 4 \sin^2\left({\sigma_k +  t_l \over 2} \right)
\right)^{\omega_k\over \sqrt{2}} \prod_{r<s}^m  \Big( 4
\sin^2\Big({t_r- t_s \over 2} \Big)\Big) \,.
}}
Again, for given $\omega_i$, at most one term in the sum over $m$ 
contributes. We recognise this formula as the perturbative open string
expansion, where one calculates Neumann boundary correlators involving 
the $n$ fields labelled by $\omega_i$, as well as an arbitrary number
of perturbing fields $J^+$. [Indeed, the correlation functions that 
appear in \anper\ are precisely the standard Neumann boundary 
Greens functions.] This formula was, for example, used as a starting   
point in \vijay.

The above amplitudes \finaltwo\ and \anper\ are formally divergent. As
is clear from the arguments of the appendix, they are to be understood
as the limit of (A.22) (or its analogue for the case of the $n$-point
amplitude) as $\tau_i\rightarrow 0$. This gives a prescription
for how to evaluate them. 
[In the above we have assumed 
that this limit exists. If this is not the case, the resulting
expressions need to be regularised further by subtracting off the
contributions from open string operators with smaller conformal weight
that appear when the bulk operators approach the boundary.]

\subsec{Cylinder partition function}

For the Euclidean theory the boundary state formalism can also be used
to calculate the cylinder diagram and thus determine the open string
spectrum. This is possible since in the boundary state of the
Euclidean theory
\eqn\eight{
|\!| B \rangle\!\rangle = \sum_{j} \sum_{m\geq 0}^{j} 
{j+m \choose 2m} (-\lambda)^{2m} \, 
|j,m,m\rangle\!\rangle }
the Ishibashi states $|j,m,m\rangle\!\rangle$ are
pairwise orthogonal. The  calculation is most easily done using a
trick that was first described in \GaberdielZQ. As was mentioned
before, the boundary state $|\!|B\rangle\!\rangle$ 
is the projection of the usual ${\rm SL}(2,\Cop)$ boundary state  
\eqn\usual{
|\!| g \rangle\!\rangle = 
\sum_{j,m,n} D^{j}_{-m,n}(g) \, |j,m,n\rangle\!\rangle }
corresponding to the group element $g=g_\lambda$ \rsixa,
where one projects onto states for which $m=n$. In terms of the matrix
elements of ${\rm SL}(2,\Cop)$ this projection can be described as 
\eqn\two{
{1\over \pi} \int_0^\pi d\theta \, D^j_{-m,n} \left( 
\pmatrix{e^{i\theta} & 0 \cr 0 & e^{-i\theta}} \, g_\lambda \, 
\pmatrix{e^{i\theta} & 0 \cr 0 & e^{-i\theta}} \right) \,.}
Up to an overall normalisation the cylinder diagram between two 
boundary states with $\lambda_1$ and $\lambda_2$ is therefore 
\eqn\three{
{1\over \pi} \int_0^\pi d\theta 
\sum_{j\in{1\over 2} {\bf Z}_+} 
\hbox{Tr}_j \left[
\pmatrix{e^{i\theta} & 0 \cr 0 & e^{-i\theta}} \, g^{-1}_{\lambda_1}
\pmatrix{e^{i\theta} & 0 \cr 0 & e^{-i\theta}} \, g_{\lambda_2}
\right]\, \chi_{j^2}(q) \,,}
where $\chi_{j^2}(q)$ is given by 
\eqn\threea{
\chi_{j^2}(q) = \vartheta_{\sqrt{2} j}(q) - \vartheta_{\sqrt{2}(j+1)}(q)
\,,}
with
\eqn\threeb{
\vartheta_s(q)  = {q^{{1\over 2} s^2} \over \eta(q)}\,.}
For each $\theta$, the trace in \three\ is simply 
${\sin((2j+1)\alpha) \over \sin\alpha}$ for some
$\alpha\equiv\alpha(\lambda_1,\lambda_2,\theta)$. In order to
determine $\alpha$, consider the $j=1/2$ representation for which we
get 
\eqn\fourcy{\eqalign{
2\cos\alpha &= \hbox{tr} \left[ 
\pmatrix{e^{i\theta} & 0 \cr 0 & e^{-i\theta}} \,
\pmatrix{1 & \lambda_1 \cr 0 & 1 } \,
\pmatrix{e^{i\theta} & 0 \cr 0 & e^{-i\theta}} \,
\pmatrix{1 & -\lambda_2 \cr 0 & 1 } \right] \cr
& = 2 \cos(2\theta)\,,}}
from which it follows that $\alpha=2\theta$. Thus the open string
spectrum is 
\eqn\fivecy{
{1\over \eta(\tilde{q})} \sum_{n\in {\bf Z}} \int_0^1 ds \, 
\tilde{q}^{(n+s)^2} = 
{1\over\eta(\tilde{q})} \int_{-\infty}^{\infty} ds \,
\tilde{q}^{s^2}\,.} 
In particular, it follows that \fourcy\ is  independent of $\lambda_1$
and $\lambda_2$. Furthermore, the conformal weights of the open string
states are real and non-negative (and not all equal to zero), 
corresponding to arbitrary real momenta in the open string.
This is then nicely in accord with our calculation for the
open string two-point function where the boundary vertex operators 
that are obtained as one takes the bulk operators to the boundary
are also labelled by arbitrary real momenta.

\newsec{Continuation to timelike signature}

As we have mentioned in the introduction, the Euclidean boundary
conformal field theory we have discussed up to now is related by an
analytic continuation $X\mapsto -i X_0$ to the timelike boundary
conformal field theory describing the half S-brane. This Wick rotation
is however rather subtle, and some of its aspects have not been
satisfactorily understood. According to Sen's original  
proposal\foot{We thank Ashoke Sen for a detailed discussion about
this issue.}, the Wick rotation should be performed for
physically meaningful quantities in the corresponding string theory;  
this can be justified because this analytic continuation maps string
solutions of the Euclidean theory to string solutions of the timelike 
theory. However, it is less clear to what extent this analytic
continuation can also be performed at the level of the underlying
conformal field theory.

The prime example for which Sen's proposal can be applied concerns the
`profile' $f(x)$ of the D-brane. Physically the function $f(x)$ can be
interpreted as determining  the transverse components of the stress
energy tensor  $T_{ij}= - {\cal T}_p \, f(x)\, \delta_{ij}$ of the
brane.  As was shown in \refs{\SenNU} it can be determined from the
bulk 1-point functions, which for the above D-brane in the Euclidean
theory are given by  
\eqn\sennull{
\tilde{f}(p)\equiv
\langle 0|\, e^{-ipx} \, |\!|g_\lambda\rangle\!\rangle = 
(-\lambda)^{{p\over \sqrt{2}}}\, \delta_{p,\sqrt{2}{\Bbb N}_0}  \,.}
The profile of the D-brane is then simply the Fourier transform of
$\tilde{f}(p)$, 
\eqn\seneq{
f(x) = {1\over 1 + \lambda \, e^{-i\sqrt{2} x}} \,.}
This function can be Wick-rotated, leading to 
\eqn\seneqp{
g(x) = {1\over 1 + \lambda\, e^{\sqrt{2} x}} \,.}
This then describes the profile of the timelike D-brane; its Fourier
coefficients \refs{\GutperleXF} 
\eqn\seneqpp{
\tilde{g}(p) =   - {\pi i\over \sqrt{2}} \,
e^{-{i\over \sqrt{2}} p \log(\lambda)} 
{1\over \sinh(\pi p /\sqrt{2})} }
can then be identified with the bulk 1-point functions of the 
timelike theory. This procedure thus defines a method for the
`analytic continuation' of the bulk 1-point function \sennull\ to 
\seneqpp. This example also shows that this analytic continuation may
be quite subtle from the point of view of the underlying conformal
field theory.

Actually, the knowledge of \seneqpp\ appears to be sufficient to
determine the boundary state in the timelike theory
uniquely.\foot{This was already pointed out in \refs{\deBoerHD}.}  For
a timelike boson, the states of a given real momentum $p\ne 0$ form
already an irreducible Virasoro representation, and thus there is a
Virasoro Ishibashi state $|p\rangle\!\rangle$ for each such
momentum. [This Ishibashi state is of `Dirichlet type', {\it i.e.} it
is explicitly 
given by 
\eqn\intbob{ |p\rangle\!\rangle = 
\exp\Big( -\sum_{n>0} {1\over n} \alpha^0_{-n} \bar{\alpha}^0_{-n} \Big) \,
|p\rangle\,,}
where $|p\rangle$ is the momentum ground state with momentum $p$, and
we have denoted the modes of the timelike boson by $\alpha^0_n$ and
$\tilde{\alpha}^0_n$.] 
The only subtlety concerns the sector with $p=0$,
since there are infinitely many Virasoro Ishibashi states with $p=0$
that are labelled by a positive integer. The bulk 1-point functions we
have discussed above determine already uniquely the coefficients in
front of the Ishibashi states corresponding to real momenta with $p\ne
0$; if we ignore the subtlety concerning the sector with $p=0$, the
boundary state must therefore be of the `integrated form'
\refs{\GutperleXF,\deBoerHD} 
\eqn\intbost{|\!|B\rangle\!\rangle^{(T)} 
= \int dp\, {\lambda^{-ip/ \sqrt{2}} \over
\sinh(\pi p/ \sqrt{2})}\,\, | p \rangle\!\rangle\,.}
However, this argument misses non-renormalisable components of the
boundary state, {\it i.e.} contributions from Ishibashi states that
come from sectors where the momentum is not real. These 
non-renormalisable parts are probably important for the 
calculation of higher point functions \vijay, the consistency of 
the bulk-boundary OPE and the presence of infinitely many conserved
charges in two dimensional string theory \SenYV. They may also be
needed in order to make sense of the cylinder diagram of two such
boundary states \KLMS. 

At any rate, it would be interesting to determine physical quantities
of the timelike theory using the above analytic continuation from the
Euclidean description. This should be possible not just for the bulk
one-point functions, but also for more complicated correlators. It
would then be interesting to compare these results with what can be
directly computed from the boundary state \intbost, using for example
the techniques described in this paper.   
\medskip

Finally, we should mention that while the boundary deformation \five\
in the Euclidean theory is not hermitian, this does not mean that it
is without any physical significance. Non-hermitian quantum mechanical
systems and defect conformal field theories have played an important 
role in condensed matter theory (for a recent example see
\NelsonA). Thus the results for the correlation functions of the
Euclidean theory are also interesting in their own right.

\vskip12pt

\noindent{\bf Acknowledgements}
\vskip4pt

We thank Costas Bachas, Stefan Fredenhagen, Per Kraus, Volker
Schomerus, Ashoke Sen and in particular Andreas Recknagel for helpful
discussions. This work was begun while we were attending the 36th
International Symposium Ahrenshoop, and we thank the organisers for
organising a stimulating meeting. The research of MRG is supported in
part by the Swiss National Science Foundation. The work of MG is
supported in part by NSF grant PHY-0245096. Any opinions, 
findings and conclusions expressed in this material are those of the
authors and do not necessarily reflect the views of the National
Science Foundation.

\appendix{A}{Boundary correlators as limits of bulk correlators}

The same methods which were used in section 2 to calculate boundary
correlators are also useful to calculate correlation functions 
involving bulk operators. In the context of the rolling tachyon
theory, such bulk correlators are related to corrections of the decay
of the unstable D-brane into closed strings. In this appendix we focus
on the relation between the bulk and the boundary vertex operators. In
particular we want to show that the open string normal ordering 
\sone\ can be derived by taking the 
leading term as the bulk vertex operators approach the 
boundary. In the following we shall first discuss the Neumann case
before applying the relevant ideas to the case with a non-trivial
boundary deformation. 

\subsec{The simple Neumann case}

It is instructive to consider first the simple case of a standard
Neumann boundary condition. The two point function ${\cal B}_2^N$ of
two bulk vertex operators is given by
\eqn\apone{
{\cal B}_2^N = \langle 0| \, : e^{i\omega_1 X(\sigma_1,\tau_1)}: \, 
: e^{i\omega_2 X(\sigma_2,\tau_2)}: \,
|\!| N \rangle\!\rangle \,,}
where the normal ordering is now the standard closed string normal
ordering prescription, {\it i.e.}
\eqn\aptwo{
: e^{i\omega X(\sigma,\tau)}: = e^{i\omega X_<(\sigma,\tau)} \,
e^{i\omega x_0}\, e^{{i\over 2} \omega p \tau} \, e^{i\omega
X_>(\sigma,\tau)} \,,}
with
\eqn\apthree{
X_{>}(\sigma,\tau) = {i\over 2} \sum_{n>0} 
{1\over n} e^{-in\tau} \left(\alpha_n e^{-in\sigma} + 
\tilde\alpha_n e^{in\sigma} \right) \,,}
and similarly for 
$X_<(\sigma,\tau)$. Here $\tau$ is purely imaginary, and lies in fact
in the lower half-plane.
Since the negative modes
annihilate the vacuum to the left, we can eliminate them by commuting
them through to the left. Applying the BCH formula we then get
\eqn\apfour{\eqalign{
{\cal B}_2^N & =  \left[ 4
e^{-i(\tau_1-\tau_2)} \, 
\sin\left({\sigma_1+\tau_1 -(\sigma_2+\tau_2) \over 2}\right) \, 
\sin\left({\sigma_1-\tau_1 -(\sigma_2-\tau_2)\over 2}\right) 
\right]^{{\omega_1 \omega_2\over 4}} \, \cr
 & \qquad \qquad \qquad \qquad \times 
\langle 0| \,e^{i\omega_1 x_0}\, e^{{i\over 2} \omega_1 p \tau_1} \, 
e^{i\omega_2 x_0}\, e^{{i\over2} \omega_2 p \tau_2} \,
e^{i\omega_1 X_>(\sigma_1,\tau_1)}\, 
e^{i\omega_2 X_>(\sigma_2,\tau_2)} \, 
|\!| N \rangle\!\rangle \,,}}
where we have used that 
\eqn\apfive{\eqalign{
[X_>(\sigma_1,\tau_1),X_<(\sigma_2,\tau_2)] 
& =   - {1\over 4} 
\log\left[ 4
e^{-i(\tau_1-\tau_2)} \, 
\sin\left({\sigma_1+\tau_1 -(\sigma_2+\tau_2)\over 2}\right) \,
\right.  \cr
&  \qquad \qquad \qquad \qquad\qquad \qquad \left. \times
\sin\left({\sigma_1-\tau_1 -(\sigma_2-\tau_2)\over 2}\right) \right] 
\,.}}
Next we move the $p$-mode to the right, using that 
$p\, |\!| N \rangle\!\rangle = 0$, and obtain
\eqn\apsix{\eqalign{
{\cal B}_2^N & =  e^{{i\over 2}\omega_1\,\omega_2 \tau_1}\, \left[ 4
e^{-i(\tau_1-\tau_2)} \, 
\sin\left({\sigma_1+\tau_1 -(\sigma_2+\tau_2) \over 2}\right) \, 
\sin\left({\sigma_1-\tau_1 -(\sigma_2-\tau_2)\over 2}\right) 
\right]^{{\omega_1 \omega_2\over 4}} \, \cr
&  \qquad \qquad \times 
\langle 0| \,e^{i(\omega_1+\omega_2) x_0}\, 
e^{i\omega_1 X_>(\sigma_1,\tau_1)+ i\omega_2 X_>(\sigma_2,\tau_2)} \, 
|\!| N \rangle\!\rangle \,.
}}
It remains to simplify
the exponential of the positive modes. To this end we
write
\eqn\apseven{\eqalign{
e^{i\omega_1 X_>(\sigma_1,\tau_1)+ i\omega_2 X_>(\sigma_2,\tau_2)} 
& =   e^{i\omega_1 X_<(\sigma_1,-\tau_1)
 + i\omega_2  X_<(\sigma_2,-\tau_2)} \, e^{C} \,  \cr
&  \quad  \times  \,
e^{i\omega_1 (X_>(\sigma_1,\tau_1)-X_<(\sigma_1,-\tau_1))
+ i\omega_2 (X_>(\sigma_2,\tau_2)-X_<(\sigma_2,-\tau_2))}\,, 
}}
where $C$ is the commutator term (as follows from the BCH formula),
\eqn\apeight{
C = {1\over 2} \left[i\omega_1 X_>(\sigma_1,\tau_1)
+ i\omega_2 X_>(\sigma_2,\tau_2),i\omega_1 X_<(\sigma_1,-\tau_1)
 + i\omega_2  X_<(\sigma_2,-\tau_2)\right] \,.
}
The idea of this construction is that the exponential that appears to
the right in \apseven\ annihilates the Neumann boundary state,
whereas the exponential to the left annihilates the vacuum. Thus it
only remains to determine $C$, which using \apfive\  equals
\eqn\apnine{\eqalign{
C & =  {\omega_1^2\over 8} \, \log \left[
-4 e^{-2i\tau_1} \, \sin^2(\tau_1) \right] 
+ {\omega_2^2\over 8} \, \log \left[
-4 e^{-2i\tau_2} \, \sin^2(\tau_2) \right]  \cr
&  \qquad + {\omega_1 \omega_2 \over 8}
\log \left[4 e^{-i(\tau_1+\tau_2)}\,
\sin\left({\tau_1+\tau_2 + \sigma_2-\sigma_1\over 2}\right) \, 
\sin\left({-(\tau_1+\tau_2) + \sigma_2-\sigma_1\over 2}\right) \,
\right]  \cr
&  \qquad + {\omega_1 \omega_2 \over 8}
\log \left[4 e^{-i(\tau_1+\tau_2)}\,
\sin\left({\tau_1+\tau_2 + \sigma_1-\sigma_2\over 2}\right) \, 
\sin\left({-(\tau_1+\tau_2) + \sigma_1-\sigma_2\over 2}\right) \,
\right]\,. }}
The total amplitude is then equal to zero unless
$\omega_1+\omega_2=0$; if this is the case, it equals 
\eqn\apten{\eqalign{
{\cal B}_2^N & =  e^{-{i\over 2}\omega^2 \tau_1}\, \left[ 4
e^{-i(\tau_1-\tau_2)} \, 
\sin\left({\sigma_1+\tau_1 -(\sigma_2+\tau_2) \over 2}\right) \, 
\sin\left({\sigma_1-\tau_1 -(\sigma_2-\tau_2)\over 2}\right) 
\right]^{-{\omega^2\over 4}} \,  \cr
&  \qquad \times 
\left[-4 e^{-2i\tau_1} \, \sin^2(\tau_1) 
\right]^{\omega^2\over 8} \,
\left[-4 e^{-2i\tau_2} \, \sin^2(\tau_2) 
\right]^{\omega^2\over 8}  \cr
&  \qquad \times
\left[4 e^{-i(\tau_1+\tau_2)}\,
\sin\left({\tau_1+\tau_2 + \sigma_2-\sigma_1\over 2}\right) \, 
\sin\left({-(\tau_1+\tau_2) + \sigma_2-\sigma_1\over 2}\right) \,
\right]^{-{\omega^2 \over 8}}  \cr
&  \qquad \times
\left[4 e^{-i(\tau_1+\tau_2)}\,
\sin\left({\tau_1+\tau_2 + \sigma_1-\sigma_2\over 2}\right) \, 
\sin\left({-(\tau_1+\tau_2) + \sigma_1-\sigma_2\over 2}\right) \,
\right]^{-{\omega^2 \over 8}} 
\,,}}
where $\omega=\omega_1=-\omega_2$, say.
The bulk vertex operators approach the boundary when 
$\tau_1,\tau_2\rightarrow 0$. In this limit the amplitude goes to zero
as
\eqn\apeleven{
\sin(\tau_1)^{{\omega_1^2 \over 4}} \, 
\sin(\tau_2)^{{\omega_2^2 \over 4}} \,.
}
If we divide by this term in order to obtain the leading behavior in
this limit, we obtain
\eqn\aptwelve{
{\cal B}_2^N  \simeq \left[4 \sin^2 \left({\sigma_1-\sigma_2 \over 2} 
\right) \right]^{-{\omega^2\over 2}} \,,}
which therefore does indeed agree with \sfive. Furthermore, using the
general structure of the bulk-boundary OPE we can deduce the conformal
weight of the boundary field. Indeed, the exponent of \aptwelve\ can
be identified with 
\eqn\apthirteen{
-{\omega^2 \over 4} = h_\omega + \bar h_\omega - h^{op}_\omega\,,}
where $h_\omega$ and $\bar h_\omega$ are the left- and right-moving
conformal weights of the bulk operator, while $h^{op}_\omega$ is the 
conformal weight of the corresponding boundary operator. With our
conventions, 
\eqn\apfifteen{
h_\omega = \bar h_\omega = {\omega^2 \over 8}\,,
}
thus leading to 
\eqn\apsixteen{
h^{op}_\omega = {\omega^2 \over 2} \,,}
which is indeed in agreement with the conformal behavior of
\aptwelve. 

\subsec{Non-trivial boundary condensate}

The calculation of the bulk two point function ${\cal B}_2$, where
we replace the Neumann boundary state $|\!|N\rangle\!\rangle$ by
$|\!|B\rangle\!\rangle$, is almost identical. 
The only difference is that now the operator $\hat Y$ in 
\eqn\apsevt{
\exp\big(\hat{Y}\big) = \exp\big( i\omega_1 
(X_>(\sigma_1,\tau_1)-X_<(\sigma_1,-\tau_1))
+ i \omega_2 (X_>(\sigma_2,\tau_2)-X_<(\sigma_2,-\tau_2))\big)}
does not annihilate $|\!|B\rangle\!\rangle$ any more. However, it
still has an expansion of the form
\eqn\apeitt{
\hat{Y} = -{1\over 2} \sum_{n\ne 0} {1\over |n|} \, 
\left(\alpha_n + \tilde\alpha_{-n} \right) \, \hat\gamma_n \,,}
where $\hat\gamma_n$ is now
\eqn\apninet{
\hat\gamma_n = \omega_1 e^{-i|n|\tau_1} e^{-in\sigma_1} + 
\omega_2 e^{-i|n|\tau_2} e^{-in\sigma_2} \,.}
In particular, we can  use essentially the same calculation
as in \seight\ -- \sthirteen, the only minor difference being that now
\eqn\apptwent{\eqalign{
\sum_{n\ne 0} {\hat\gamma_n \over |n|} e^{- i t n}
& = 
- \omega_1 \log\left(1 - e^{-i(\sigma_1 + \tau_1+  t)} \right)
- \omega_1 \log\left(1 - e^{i(\sigma_1-\tau_1 +  t)} \right)  \cr
&  \qquad 
- \omega_2 \log\left(1 - e^{-i(\sigma_2+\tau_2 +  t)} \right)
- \omega_2 \log\left(1 - e^{i(\sigma_2 -\tau_2+   t)} \right) \,.
}}
The relevant $Z$ is therefore now 
\eqn\aptwone{\eqalign{
Z & =  - \lambda \,  \int_0^{2\pi} {dt\over 2\pi}\, J^+(e^{- i t}) 
\exp\left( - {1\over \sqrt{2}} \, 
\sum_{n\ne 0} {\hat\gamma_n \over |n|} e^{- i t n} \right)  \cr
& =  
-\lambda \,  \int_0^{2\pi} {dt\over 2\pi}\,  J^+(e^{- i t})  
\left(1 - e^{-i(\sigma_1 +\tau_1 +  t)} 
\right)^{{\omega_1\over \sqrt{2}}} \,
\left(1 - e^{i(\sigma_1 -\tau_1 +  t)} 
\right)^{{\omega_1\over \sqrt{2}}}  \cr
&  \qquad \qquad \qquad \qquad \qquad \qquad \times
\left(1 - e^{-i(\sigma_2 +\tau_2 +  t)} 
\right)^{{\omega_2\over \sqrt{2}}} \,
\left(1 - e^{i(\sigma_2 -\tau_2 +  t)} 
\right)^{{\omega_2\over \sqrt{2}}}  \,.}}
Thus the total amplitude  can be expressed as
\eqn\aptwtwo{\eqalign{
{\cal B}_2 &=   \langle 0| \, : e^{i\omega_1 X(\sigma_1,\tau_1)}: \, 
: e^{i\omega_2 X(\sigma_2,\tau_2)}: \,
|\!| B \rangle\!\rangle  \cr 
&= {\cal B}_2^N\, \langle 0 | e^{i(\omega_1+\omega_2)x_0} e^Z | e
\rangle\!\rangle\,,}} 
where ${\cal B}_2^N$ is the two point function with Neumann boundary
condition \apten. Expansion as a power series in $\lambda$ then gives 
\eqn\aptwthree{\eqalign{
{\cal B}_2  & =  {\cal B}_2^N \, \sum_{m=0}^{\infty}
{(-\lambda)^m\over m!}
\delta(\omega_1+\omega_2+ \sqrt{2}m) \prod_{i=1}^m  \int_0^{2\pi}
{dt_i\over 2\pi}\, 
\prod_{k<l}^{m} \Big( 4 \sin^2\Big({t_k- t_l\over 2}\Big) \Big)  \cr
&\qquad \qquad \times  \prod_{l=1}^m \left[ 
 \left(1 - e^{-i(\sigma_1 +\tau_1 +  t_l)} 
\right)^{{\omega_1\over \sqrt{2}}} \,
\left(1 - e^{i(\sigma_1 -\tau_1 +  t_l)} 
\right)^{{\omega_1\over \sqrt{2}}}\right. \cr
& \left. \qquad \qquad \qquad \times    
\left(1 - e^{-i(\sigma_2 +\tau_2 +  t_l)} 
\right)^{{\omega_2\over \sqrt{2}}}
\left(1 - e^{i(\sigma_2 -\tau_2 +  t_l)} 
\right)^{{\omega_2\over \sqrt{2}}} \right]\,.
}}
Since the $\tau_i$ are purely imaginary (and lie in the lower
half-plane), these integrals converge. In the limit
$\tau_1,\tau_2\rightarrow 0$ the factor ${\cal B}_2^N$ tends again to
zero as in \apeleven; if the integrals in \aptwthree\  converge to a
finite answer in this 
limit\foot{If this is not the case, then open string vertex operators
with smaller conformal weight appear in the limit in which the bulk
operators approach the boundary. Their contributions must then be
subtracted off, {\it i.e.} the divergent integrals in \aptwthree\ must
then be further regularised.}, 
${\cal B}_2$ goes to zero as
\eqn\aptwfour{
\sin(\tau_1)^{{\omega_1^2 \over 4}} \, 
\sin(\tau_2)^{{\omega_2^2 \over 4}} \,.}
This implies, by the same reasoning as before, that we need to divide 
through by this factor, in order to isolate the leading
contribution. The relevant limit then gives 
the 2-point boundary correlator of two non-trivial vertex
operators with conformal weight $h_\omega^{op}=\omega^2/2$. 
On the other hand, it is clear by construction that in this limit 
${\cal B}_2$ becomes the function ${\cal A}_2$ we have calculated in
section~2.

\listrefs

\end